# Two-dimensional germanium islands with Dirac signature on Ag$_2$Ge surface alloy

Jiaqi Deng[1], Gulnigar Ablat[1], Yumu Yang[1], Xiaoshuai Fu[1], Qilong Wu[1], Ping Li[2*], Li Zhang[1], Ali Safaei[3], Lijie Zhang[1*], and Zhihui Qin[1]

1. Key Laboratory for Micro/Nano Optoelectronic Devices of Ministry of Education & Hunan Provincial Key Laboratory of Low-Dimensional Structural Physics and Devices, School of Physics and Electronics, Hunan University, Changsha, 410082, China
2. Center for Spintronics and Quantum Systems, State Key Laboratory for Mechanical Behavior of Materials, School of Materials Science and Engineering, Xi'an Jiaotong University, Xi'an, Shaanxi, 710049, China
3. Metallurgy and Materials Engineering Department, University of Gonabad, Khorasan Razavi, Iran

E-mail: Ping Li: pli@xjtu.edu.cn; Lijie Zhang: lijiezhang@hnu.edu.cn

## Abstract

Two-dimensional (2D) Dirac materials have attracted intense research efforts due to their promise for applications ranging from field-effect transistors and low-power electronics to fault-tolerant quantum computation. One key challenge is to fabricate 2D Dirac materials hosting Dirac electrons. Here, monolayer germanene is successfully fabricated on a Ag$_2$Ge surface alloy. Scanning tunneling spectroscopy measurements revealed a linear energy dispersion relation. The latter was supported by density functional theory calculations. These results demonstrate that monolayer germanene can be realistically fabricated on a Ag$_2$Ge surface alloy. The finding opens the door to exploration and study of 2D Dirac material physics and device applications.

1. **Introduction**

Two-dimensional (2D) materials have attracted a lot of attention due to their exotic properties and appealing applications since the successful isolation of graphene.[1,2] Germanene, the germanium analog of graphene, shares many properties with graphene.[3,4] The electrons in germanene are predicted to host Dirac properties such as linear dispersing energy bands near the K points of the Brillouin zone.[5,6] Furthermore, germanene has a much larger spin-orbit gap than graphene,[7] making it suitable for different applications. In order to exploit the unique properties of 2D materials for technological applications, it is of utmost importance to master the art of controllable growth of such materials with clearly understood growth mechanism and parameters. Despite many previous attempts on different substrates,[8-14] its growth and subsequent characterization remain inconclusive.

Theoretical calculation demonstrated that germanene can adhere to Ag(111) via electrostatic interactions.[15] However, Dávila *et al.[9]* proposed that germanene will not be grown on Ag(111) due to the lattice mismatch and the possibility of forming $Ag_2Ge$ surface alloy.[16] Moreover, prior to the emergence of germanene, germanium had also been found to form a $Ag_2Ge$ surface alloy on Ag(111) surface when 1/3 monolayer Ge deposited on the substrate.[16,17] Recently, Lin *et al.* found two phases of germanene on Ag(111) substrate, the so-called striped phase and quasi-freestanding phase,[18] attributed to the commensurability of the adlayer with the substrate. Zhuang *et al.* reported on the successful growth of germanene on a thin film of germanium that was deposited on a Ag(111) substrate,[19] though earlier reports had claimed that germanium deposition on Ag(111) leads to the formation of only a crystalline film rather than anything else.[20] Quite recently, Zhang *et al.* reported the growth of germanium on Ag(111) at a specific temperature range of 380-430 K. The authors extrapolated the obtained disorder hexagonal structure after higher amounts Ge deposition contains Ag as well.[21] However, it still lacks direct evidence since the STM do not have an element-specific capability. Furthermore, electronic properties measurement of the formed adlayer, for instance scanning tunneling spectroscopy data, were lacking in their study. Overall, the growth of germanium on Ag(111) is still unclear. A scrutinization of the growth of germanium on Ag(111) will be helpful in understanding the formation of germanene.

In this work, we employ scanning tunneling microscopy (STM) and first-principle density functional theory (DFT) calculations to investigate the deposition and growth of germanium on Ag(111) surface. At a low deposition flux, irregular-shaped 2D islands of germanium are formed on the substrate. Upon increasing the deposition flux, the irregular-shaped islands convert to compact 2D islands. These germanium islands are grown in 2D mode, and their differential conductivity is found to be V-shaped around the Fermi level, which is a hallmark of germanene, a 2D Dirac material.[22,23] These 2D germanium islands have always been found to grow exclusively on the previously formed domain of a different phase that we believe is the $Ag_2Ge$ surface alloy. Meanwhile, our DFT calculations show that germanene on a bilayer $Ag_2Ge$ surface alloys with a V-shaped profile of density of sates near the Fermi level. Therefore, our findings suggest that we have successfully synthesized germanene on a bilayer of

Ag$_2$Ge surface alloy supported by Ag(111) surface. The present work suggests that the Ag$_2$Ge surface alloy could be the precursor phase for the formation of germanene.

**2. Methods**

The experiments were carried out at a home-made ultrahigh vacuum (UHV) chamber (<1.0×10$^{-10}$ millibar) equipped with an STM (Unisoku Co., Ltd.). The substrate was a single-crystal Ag(111), and was cleaned by several cycles of Ar$^+$ ion sputtering (1 keV) with subsequent annealing in UHV. The deposition of germanium was done by heating high-purity Ge seeds (Alfa Aesar Puratronic 99.9999%) in a commercial Knudsen cell mounted on the UHV chamber. Subsequently, we annealed the sample mildly (around 450 K) for 5 minutes. The deposition flux was roughly calibrated by the coverage on the substrate. All the STM measurements were made at room temperature. A standard lock-in amplifier was employed for the differential conductivity (dI/dV) measurements by a modulating signal of 50 mV and 773 Hz. The dI/dV spectra shown in this work are all point measurements.

Our DFT-based first-principles calculations were implemented using the *Vienna ab initio simulation package* (VASP).[24,25] The electron exchange correlation functional was described by the generalized gradient approximation of the Perdew-Burke-Ernzerhof functional.[26] The plane-wave basis set with a kinetic energy cutoff of 400 eV was used. The geometry optimization for obtaining the final structures was based on minimizing the remaining Hellmann-Feynman forces to less than 0.01 eV/Å. 9×9×1 (12×12×1) k-meshes were adopted for the structural optimization and the self-consistent calculations. To avoid unnecessary interactions between the monolayer, the vacuum layer was set to 20 Å.

## 3. Results and discussion

An atomically clean Ag(111) substrate is required for the deposition of germanium. Figure 1a shows a large-scale STM image of a clean Ag(111) substrate. The measured step height of 0.22 nm agrees well with the inter-planar distance of (111) planes of a face-centered cubic (FCC) crystal with the lattice constant of 0.289 nm.[27] The empty-state STM image in Figure 1b reveals that the periodicity of the surface texture is about 0.28 nm which fits quite well with the lattice constant of the Ag(111)-(1×1) structure. The corresponding fast Fourier transform (FFT) depicted in the inset image of Figure 1b agrees well with the high-resolution STM image.

Figure 1c shows the morphology of the surface with the deposition of germanium for 15 min with the rate of about 0.022 ML/min and an annealing subsequently. Irregular-shaped domains with two distinct contrasts are visible in the figure. It has been reported that the substitution of very small amount of Ge and Ag to form dark spots due to the smaller radius of Ge than Ag.[17] Similarly, the dark area here are the enlarged spots with increasing amount of substitution of Ge with top layer Ag. And, these dark areas act as nucleation centers for the growth of 2D Ge islands.[28] It explains why these germanium layer are all located on top of the dark region. The branches of the irregular-shaped islands are of monoatomic height with constant thickness, which is supported by the height-profile shown in the inset of Figure 1c. Figure 1d shows a zoom-in STM image of irregular Ge structure revealing that germanium on Ag(111) at low coverage grows in the form of 2D instead of three-dimensional (3D) islands. However, these irregular Ge clusters are not likely crystalline structure hindering the atomic resolution image resolved by STM.

To further study the growth of germanium on Ag(111), we increased the deposition flux to 0.062 ML/min, *i.e.* to about three times larger than that of the previous deposition, keeping all other parameters unchanged. The result of this deposition has been given in Figure 2a. Three different contrast levels are visible in the figure: bright domains, less-bright areas, and dark domains. Figure 2b gives two different periodicities of $0.48 \pm 0.02$ nm and $0.28 \pm 0.02$ nm which agree with the lattice constants of $Ag_2Ge$ surface alloy[17] and bare Ag(111) surface, respectively. Furthermore, the dI/dV spectrum on the less-bright areas shown in the Figure 2c exhibits a typical feature that is consistent with the known surface state onset below the Fermi level of the Ag(111) surface.[29,30] The energy shift of the typical peak of the dI/dV curve is about 50 meV, due to thermal drift.[30] Figure 2d gives the dI/dV spectrum of the dark-contrast domains of Figure 2a—the metallic character is clearly seen in this spectrum. Therefore, we identify the dark fractal domains in Figure 2a as the $Ag_2Ge$ surface alloy, and the less-bright areas of Figure 2a as the remaining parts of the pristine Ag(111) surface. Later on, we provide more evidence that the brightest areas of Figure 2a are germanium islands.

It should be noted here that the bright and dark domains in Figure 2a are compact as opposed to the corresponding irregular domains found in Figure 1c. This shape transition of the Ge domains and the $Ag_2Ge$ domains—the brightest and dark domains,

respectively—in Figure 2a in comparison to Figure 1c might indicate that the growth of the 2D islands of Ge on Ag(111) is a reaction-limited aggregation process rather than a diffusion-limited aggregation.[31] Such 2D islands transition was first observed for the growth of Ge islands on a Pb-covered Ag(111) substrate.[31] One may think that the $Ag_2Ge$ surface alloy acts as a surfactant for the growth of 2D islands of germanium.

A large-scale image of the brightest areas of Figure 2a has been given in Figure 3a. The apparent height of the bright-contrast island shown in Figure 3a is about 0.25 nm (see the inset of Figure 3a), consistent with the height of a single-layer germanium. Carefully inspecting the STM images, we observe that these 2D islands of bright contrast are indeed always on top of the dark domains that we have already identified as the $Ag_2Ge$ surface alloy; that is, these bright 2D domains always grow on top of the domains of the $Ag_2Ge$ surface alloy—the dark domains—rather than directly on the pristine Ag(111) substrate. Upon further deposition of germanium, those germanium atoms that had taken the positions of silver atoms at the top surface layer can act as nucleation centers for the growth of 2D islands of germanium—similar to what has been observed for the growth of antimonene on $SbAg_2$ surface alloy covered Ag(111),[32] and stanene/$Ag_2Sn$ on Ag(111).[33] It explains why all these germanium islands are located on top of the dark regions, *i.e.* on the $Ag_2Ge$ phase. Unfortunately, atomic resolution honeycomb structure of germanene has not achieved here. The dI/dV of these bright domains, which is proportional to their local density of states, exhibits a well-defined V-shape (see Figure 3b). Such a V-shaped curve of density of states is one of the hallmarks of a 2D Dirac system, that is, revealing a signature of linear energy dispersion relation.[22,23] It also excludes the possibility that we are dealing with the growth of regular or amorphous germanium. The residual conductivity at the zero bias in Figure 3b reflects the metallic character of the material that is beneath the germanene islands. It has been calculated that the Dirac properties of germanene can be preserved on Ag although slightly hybridized.[34] The energy range of this V-shaped curve of density of states is about 0.2 eV, which is in agreement with the prediction of germanene in previous theoretical calculation.[5] Interestingly, Zhuang *et al*.[19] found that germanene terminated on top of Ge(111) nanosheets on an Ag(111) substrate, which is, however, different from our case.

We observed that the deposition rate could substantially change the nucleation of germanium clusters. At a low deposition rate, Ge atoms can meet each other slowly; they then react with the surfactant $Ag_2Ge$ to form the Ge irregular islands. At a deposition rate three times higher than the previous one, the deposited Ge atoms are more mobile to meet and overcome the nucleation barrier to form crystalline growth mode. In a recent study, Yuhara *et al.[35]* synthesized germanene by segregating through Ag thin films, which means the Ge-Ge interaction is stronger than Ge-Ag bonding, and boosts the formation of germanene layer. A follow-up work has used a similar recipe to segregate germanene between two-dimensional material covered Ag(111) surface.[36] It is worth mention that Zhang *et al.* performed the deposition of Ge on Ag(111) and found a hexagonal structure of the adlayer with a lattice constant of 4.3 ± 0.2 Å,[21] which is, actually, quite close to the predicted freestanding germanene of 3.97Å.[5] Furthermore, the lattice constants of germanene will be slightly stretched if the buckling is lower than the freestanding condition, which had been reported in germanene/Ge$_2$Pt system.[8,23]

In order to further confirm our results, we perform first-principle DFT calculations on four different structures of germanene on monolayer, bilayer, and trilayer Ag$_2$Ge substrate. We have found that only the germanene/bilayer-Ag$_2$Ge bridge-Ag-Ag structure gives rise to a well-defined V-shaped profile of density of states in the vicinity of the Fermi level (see Figure 4a), which is in good agreement with the experimental findings. Figure 4b and 4c show, respectively, the top- and side-views of the structure. In other words, the germanene-on-monolayer-Ag$_2$Ge and the germanene-on-trilayer-Ag$_2$Ge systems never show a V-shaped curve of density of states near their Fermi levels (see Figure S1 in supplementary material). To understand this, we need to first analyze the optimized structures. Table S1 (in supplement material) gives the optimized spacing between the germanene layer and its underlying layer for different configurations. The closest calculated spacing to what we have measured experimentally (2.5 Å in Figure 3b) is 2.91 Å which happens for the germanene layer on top of a bilayer Ag$_2$Ge in the bridge-Ag-Ag configuration.

Moreover, we calculated the germanene-substrate binding energy defined as $E_b$ = ($E_{tot}$-$E_{sub}$-$E_{ger}$)/$N_{ger}$ for all possible configurations (see Table S2), with $E_{tot}$, $E_{sub}$, and $E_{ger}$ being the total energy of the germanene/substrate system, that of the substrate, and that of the germanene layer, respectively. $N_{ger}$ is the number of Ge atoms in the germanene layer. For the system of a germanene layer on a bilayer Ag$_2$Ge, the binding energy of the germanene-substrate and that of the Ag$_2$Ge layers with each other—*i.e.* the Ag$_2$Ge-

Ag$_2$Ge binding energy—are lower than these binding energies for the other two structures (the germanene-on-monolayer Ag$_2$Ge structure and germanene-on-trilayerAg$_2$Ge structure). Also, they can stably exist. Therefore, the germanene-on-bilayer Ag$_2$Ge bridge-Ag-Ag structure corresponds to our experiments.

## 4. Conclusion

In summary, we have studied the growth of germanium on Ag(111), and found the growth mode is crucially determined by the deposition rate. At low deposition rates, Ge atoms form irregular domains, whereas a layer of germanene forms when the deposition rate is three times higher. The V-shaped density of states and corresponding DFT calculations suggest that we are dealing with a germanene layer. This work is helpful to understand the controllable growth of germanene.

**Authors' contributions**

J.D., G.A., Y.Y., X.F. and Q.W. performed the STM experiments. P.L. did the DFT calculations. J.D and L.Z. took part in preparing the manuscript with the input from all co-authors. A.S. provided many technical and text editorial suggestions. L.Z. and Z.Q. coordinated the research project. All authors participated in discussing the data.

**Acknowledgements**

This work is supported by the National Natural Science Foundation of China (Grant Nos. 51972106, 11904094 and 51772087), the Strategic Priority Research Program of Chinese Academy of Sciences (Grant No. XDB30000000), and Natural Science Foundation of Hunan Province, China (Nos. 2019JJ50034 and 2019JJ50073). P. Li thanks China's Postdoctoral Science Foundation funded project (No.

2020M673364).

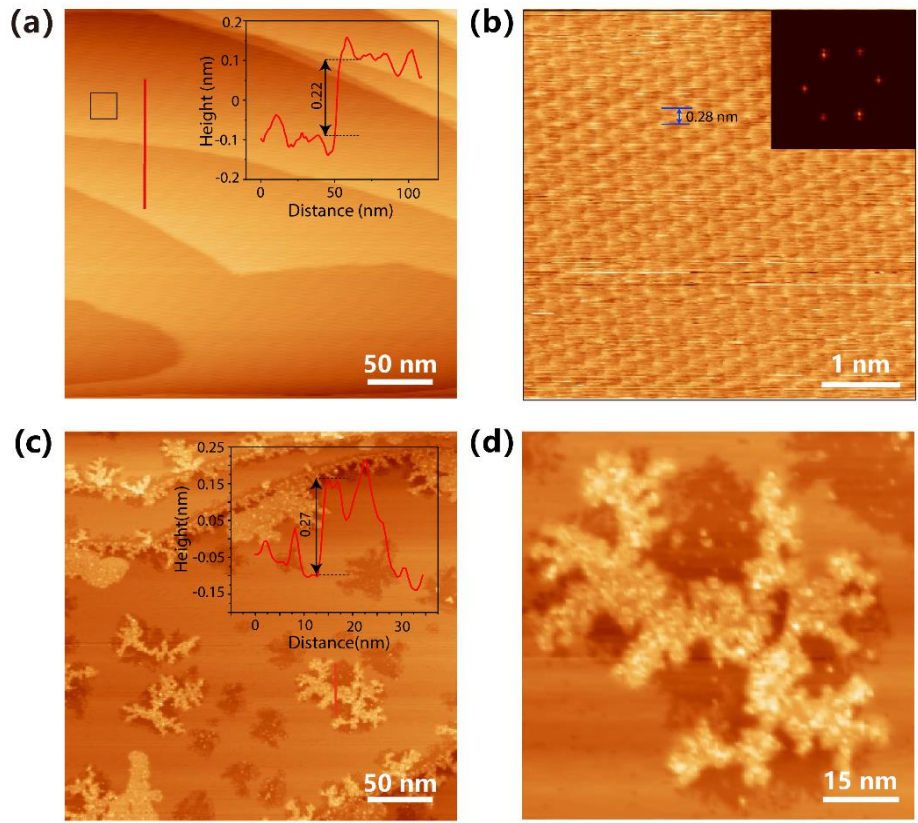

**Figure 1.** (a) Large-scale STM image of clean Ag(111) surface. The sample bias and setpoint are 1V, 100 pA. (b) Zoom-in atomic resolution STM image (200 mV, 1.2 nA) with an inset FFT. (c) Large-scale STM images of Ge-deposited Ag(111). The deposition flux was 0.02ML/min. The bright irregular islands are 2D islands of germanium, and the dark irregular islands are $Ag_2Ge$ surface alloy domains. The less-bright areas are the remaining parts of the pristine Ag(111) surface. The were -1.7 V and 10 pA. The inset shows the height profile of a germanium irregular domain. (d) Zoom-in STM image of the Ge irregular island (-1.7 V,10 pA).

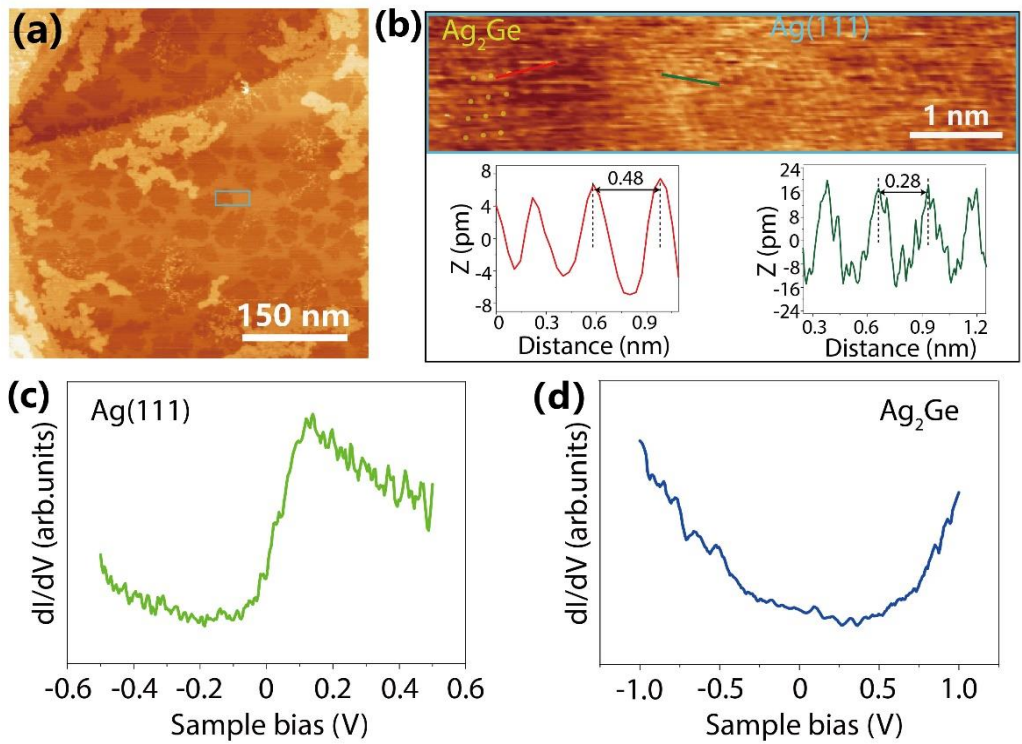

**Figure 2.** (a) Large-scale STM images of Ge-deposited Ag(111) surface. The deposition flux was 0.062 ML/min. The set points are 100 pA and 1 V. The brightest areas are 2D irregular islands made of germanium. The less-bright areas are parts of the bare Ag(111) surface that are left untouched by the deposition; and the dark-contrast areas are the irregular islands of $Ag_2Ge$ surface alloy formed upon deposition of 1/3 of monolayer of Ge at room temperature. (b) High-resolution STM image of the $Ag_2Ge$ and Ag(111) structures (800 pA, -200 mV). The Bottom-left panel and -right panels show the line profiles within the $Ag_2Ge$ and Ag(111) phases, respectively. (c) The dI/dV versus sample-bias spectrum of the Ag(111) phase, and (d) that of the $Ag_2Ge$ phase.

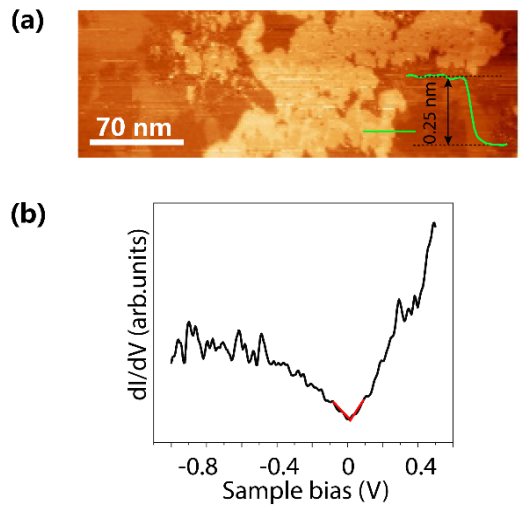

**Figure 3**. (a) A large-scale STM image of Figure 2a sample (100 pA, 1 V). The bright contrast is for germanene islands; the less-bright contrast is for the pristine Ag(111) surface. And the dark contrast is for the $Ag_2Ge$ surface alloy. The inset height profile of the germanene island through the green line segment in (a). (b) The differential conductivity record on the germanene island.

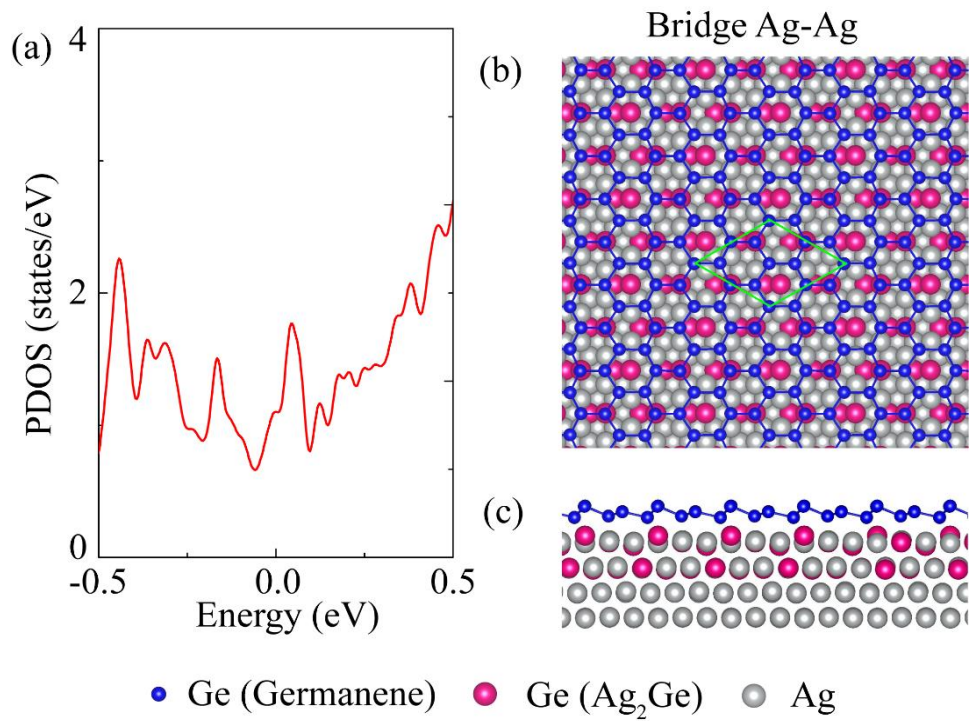

**Figure 4.** Schematic atomic structure and PDOS calculated within the DFT for bridge-Ag-Ag structure of germanene on bilayer $Ag_2Ge$ substrate. (a) The PDOS of germanene. (b) Top-view structure (c) Side-view structure. The blue, gray, and red balls are, respectively, the Ge atoms within the germanene layer, the Ag, and the Ge atoms of the $Ag_2Ge$ surface alloy layer.